\documentclass[aps,prd,twocolumn]{revtex4}

\newcommand{\be}{\begin{equation}}
\newcommand{\e}{\end{equation}}
\newcommand{\bear}{\begin{eqnarray}}
\newcommand{\ear}{\end{eqnarray}}
\newcommand{\nline}{\nonumber \\}
\newcommand{\f}{\frac}
\newcommand{\del}{\partial}
\newcommand{\dphi}{\partial_i \phi \partial^i \phi}

\begin{document}

\title{Can the clustered dark matter and the smooth dark energy arise from 
the same scalar field ?}

\author{T. Padmanabhan}
\email[]{nabhan@iucaa.ernet.in }
\homepage[]{http://www.iucaa.ernet.in/~paddy}
\author{T. Roy Choudhury}
\affiliation{IUCAA, Ganeshkhind, Pune, India 411 007}

\date{\today}

\begin{abstract}
Cosmological observations suggest the existence of two different kinds of
energy densities dominating at small ($ \lesssim 500$ Mpc) and large ($\gtrsim
1000 $ Mpc) scales. The dark matter component, which dominates at small scales,
contributes $\Omega_m \approx 0.35$ and has an equation of state $p=0$, while
the dark energy component, which dominates at large scales,  contributes
$\Omega_V \approx 0.65$ and has an equation of state $p\simeq -\rho$. It is
usual to postulate weakly interacting massive particles (WIMPs) 
for the first component and some form of scalar field
or cosmological constant for the second component. We explore the possibility
of a scalar field with a Lagrangian $L =- V(\phi) \sqrt{1 - \del^i \phi \del_i
\phi}$ acting as {\it both} clustered dark 
matter and smoother  dark energy and
having a scale-dependent equation of state.  This model predicts a relation
between the ratio  $ r = \rho_V/\rho_{\rm DM}$ of the  energy densities of the
two dark components and expansion rate $n$ of the universe 
[with $a(t) \propto
t^n$] in the form  $n = (2/3) (1+r) $. For $r \approx 2$, we get $n \approx 2$
which is consistent with observations.
\end{abstract}

\maketitle

The most conservative explanation of the current cosmological observations
will require two components of dark matter. (a) First one is a dust component
with the  equation of state  $p=0$ contributing $\Omega_m \approx 0.35$. This
component clusters gravitationally at small scales  ($l \lesssim 500$ Mpc, say)
and will be able to explain observations from galactic to supercluster scales.
(b) The second one is a negative pressure component with an 
equation of state like
$p=w\rho$  with $-1 < w < -0.5 $ contributing about $\Omega_V \approx 0.65$.
There is  some leeway in the $(p/\rho)$ of the second component but it is
certain that  $p$ is negative and $(p/\rho)$ is of order unity  (for recent
reviews, see \cite{sarkar}). The 
cosmological constant will provide $w=-1$ while
several other candidates based on scalar fields with potentials
\cite{phiindustry} will provide different values for $w$ in the acceptable
range.  By and large, component (b) is noticed only in the large scale
expansion and it does not cluster gravitationally to a significant extent.

Neither of the components (a) and (b) has laboratory evidence for its 
existence directly or indirectly. In this sense, cosmology requires invoking
the tooth fairy twice to explain the current observations. It would be  nice if
a candidate could be found which can explain the observations  at both small
and large scales (so that the tooth fairy needs to be invoked  only once). The
standard cold dark matter model of the 1980's belongs to  this class but --
unfortunately -- cannot explain the observations.  It is obvious from the
description in the first paragraph, that any such (single) candidate  must have
the capacity of leading to different equations  of state at different scales
and making a transition from $p=0$ at small scales to $p=-\rho$ (say) at large
scales. Normal particles  (that is, one-particle-excitations of  standard
quantum field theory) such as weakly interacting massive particles (WIMPs) 
will usually lead to the equation of state
$p=0$ at all scales. On the other hand, homogeneous field configurations in
scalar field models will behave like dark energy with negative pressure and
cannot cluster effectively at small scales.

In this paper we examine the possibility of whether a recently proposed
\cite{asen} candidate --- a rolling tachyon arising from string theory ---  can
explain dark matter observations {\it at both small and large scales}. 

The structure of this scalar field  can be understood by a simple analogy from
special relativity. A relativistic particle with  (one-dimensional) position
$q(t)$ and mass $m$ is described by the Lagrangian $L = -m \sqrt{1-\dot q^2}$.
It has the energy $E = m/  \sqrt{1-\dot q^2}$ and momentum $p = m \dot
q/\sqrt{1-\dot q^2} $ which are related by $E^2 = p^2 + m^2$.  As is well
known, this allows the possibility of having massless particles with finite
energy for which $E^2=p^2$. This is achieved by taking the limit of $m \to 0$
and $\dot q \to 1$, while keeping the ratio in $E = m/  \sqrt{1-\dot q^2}$
finite.  The momentum acquires a life of its own,  unconnected with the
velocity  $\dot q$, and the energy is expressed in terms of the  momentum
(rather than in terms of $\dot q$)  in the Hamiltonian formulation. We can now
construct a field theory by upgrading $q(t)$ to a field $\phi$. Relativistic
invariance now  requires $\phi $ to depend on both space and time [$\phi =
\phi(t, {\bf x})$] and $\dot q^2$ to be replaced by $\partial_i \phi \partial^i
\phi$. It is also possible now to treat the mass parameter $m$ as a function of
$\phi$, say, $V(\phi)$ thereby obtaining a field-theoretic Lagrangian $L =-
V(\phi) \sqrt{1 - \del^i \phi \del_i \phi}$. The Hamiltonian  structure of this
theory is algebraically very similar to the special  relativistic example  we
started with. In particular, the theory allows solutions in which $V\to 0$,
$\dphi \to 1$ simultaneously, keeping the energy (density) finite.  Such
solutions will have finite momentum density (analogous to a massless particle
with finite  momentum $p$) and energy density. Since the solutions can now
depend on both space and time (unlike the special relativistic example in which
$q$ depended only on time), the momentum density can be an arbitrary function
of the spatial coordinate. This provides a rich gamut of possibilities in the
context of cosmology \cite{comment}.

To examine this scenario in more detail, we will begin with the action  
which couples such a scalar field to gravity at low energies:
\be
S = \int d^4x \sqrt{-g} \left(\f{R}{16 \pi G} 
- V(\phi) \sqrt{1 - \del^i \phi \del_i \phi}\right),
\e
where $\phi$ is the  field and $V(\phi)$ is the potential. 
Though motivated from string-theoretic considerations,
we shall take this action as the starting point and investigate
its consequences without worrying about its origin. (In this spirit, we 
refer to $\phi$ as simply a {\it scalar field}, rather than as 
a tachyonic field).
 The Einstein equations are
\be
R^i_k - \f{1}{2} \delta^i_k R = 8 \pi G T^i_k
\e
where the stress tensor for the scalar  field can be written in a 
perfect fluid form
\be
T^i_k = (\rho + p) u^i u_k - p \delta^i_k
\e
with
\bear
u_k &=& \f{\del_k \phi}{\sqrt{\del^i \phi \del_i \phi}}; ~~ u_k u^k = 1
\nline
\rho &=& \f{V(\phi)}{\sqrt{1 - \del^i \phi \del_i \phi}} \nline
p &=& -V(\phi) \sqrt{1 - \del^i \phi \del_i \phi}.
\ear
The remarkable feature of this stress tensor is that it could be considered
as {\it the sum of two components ($a$) and ($b$)} described in the 
first paragraph. 
To show this explicitly, 
we  break up the density $\rho$ and the pressure $p$ 
and write them in a more suggestive form as
\be 
\rho = \rho_V  + \rho_{\rm DM}; ~~
p = p_V  + p_{\rm DM}
\e
where
\bear
\rho_{\rm DM} = \f{V(\phi) \del^i \phi \del_i \phi}
{\sqrt{1 - \del^i \phi \del_i \phi}}; &~~& p_{\rm DM} = 0
\nline
\rho_V  = V(\phi) \sqrt{1 - \del^i \phi \del_i \phi}; &~~&
p_V  = -\rho_V 
\ear
This means that the stress tensor can be thought of as made up of two components
-- one behaving like a pressure-less fluid, while the other having a negative
pressure. 

If $V(\phi)$ decreases with $\phi$ and has a minimum at $V=0$ as $\phi\to
\infty$ then it is possible to obtain pressure-less dust solutions by taking
the limit $V ~ \to ~ 0$, $\dphi ~\to ~1$ simultaneously and  keeping the energy
density finite in the $\rho_{\rm DM} $ component. If this happens globally {\it
at all scales}, then --- in this limit --- the scalar field will  behave as
pressure-less dust {\it at all scales}. In this limit $\rho_V$ will vanish.
Linear perturbation analysis shows \cite{frolov} that this component will
cluster gravitationally somewhat similar to dust-like particles. In this
scenario, the  scalar field will merely act as (yet another) candidate for dark
matter \cite{frolov}. (It may be noted that there are still some subtleties related to  clustering properties,
time-scales etc. which have to be sorted out. But we believe this is indeed possible. For example,
some of the problems related to velocities of the condensate particles can be addressed by using solutions
which are Lorentz boosted, as explained in \cite{boost}).  

It is, however, unlikely that such a scenario will be cosmologically acceptable in the
absence of  another component (b) with negative pressure described in the first
paragraph. Unless the clustering property of this scalar field is sufficiently
different from that of matter with $p=0$, we will need to still invoke a
separate component to describe cosmological observations. More generally,  if
one assumes that the field $\phi$ has the same configuration at all length
scales, then one would end up getting the same density-pressure relation
(equation of state) at all scales. However, in the  real universe, we know that
the dynamics of structure  formation and clustering at galactic scales is
dominated by the pressure-less fluid component (dark matter), while at large
scales,  the dynamics of the expansion of the universe  is governed by
spatially averaged mean density of a pressure-less component and a smooth
component with negative pressure.  In order to understand these effects, we
need to model the scalar field in such a manner that we get different equations
of state at different scales.  This is possible if we assume  that the field
$\phi$ has some sort of stochastic behaviour so that its properties at
different scales can be obtained by carrying out an averaging over the
corresponding scales.

To tackle this complicated issue, we shall define an average of any 
quantity $A[\phi(t, {\bf x})]$ over a 
length scale $R$, such that the averaged quantity describes 
the behaviour of the
field at that length scale. (This is a fairly standard practice in the study of 
structure formation; see, for example, chapter 5 of \cite{paddysfu}.)  
The average of $A(\phi)$ over a length scale $R$ is defined by smoothing 
it with a window function $W_R$. Mathematically, this is expressed as
\bear
A(R)  \equiv \langle A(\phi) \rangle_R &=& \int \f{d^3{\bf k}}{(2 \pi)^3}
A_{\bf k}(\phi) W_R({\bf k}), \nline
A_{\bf k}(\phi) &=& \int d^3{\bf x} A(\phi({\bf x})) e^{i {\bf k \cdot x}},
\ear
where $W_R({\bf k}) \propto \exp(-k^2 R^2/2)$
if the window function can be taken to be Gaussian, say. 
In this case, the behaviour at a scale $R$ will be described 
by an average potential $\bar{V}_R(\bar{\phi})$ obtained by eliminating $R$
between the average of potential $\bar{V}(R)$ and the average of field
$\bar{\phi}(R)$ when all the average quantities are obtained using the 
same window function. 
In such a description, $\phi$ will sample different parts of $V(\phi)$
at different scales and it is possible to have different equations of state
at small and large scales. 

To see how it works, consider a simple case in which the field configuration
evolves as 
\be
\phi(t, {\bf x}) = A({\bf x}) t + \f{f({\bf x})}{t^3}.
\label{phi_t}
\e
which is a simple generalization of the evolution described in  some of the
previous works (see, e.g, \cite{asen}, \cite{paddy}, \cite{gibb}, \cite{earl}).
When averaged  over a length scale $R$ we obtain an effective field
\be
\bar{\phi}(t,R) = A(R) t + \f{f(R)}{t^3}
\e
The dependence of $A(R)$ and $f(R)$ on $R$ will determine the 
behaviour of the field at different scales. 
The time dependence of the second term is appropriate 
if the effective potential at scalar $R$ behaves as 
\begin{equation}
\bar{V}_R(\bar{\phi}(t,R)) = V_0 \left(\f{\phi_0}{\bar{\phi}(t,R)}\right)^2
\end{equation}
which was considered earlier in \cite{paddy}, \cite{fein}.
For a different potential, the time dependence will be different
but in general for  $t \gg 1$,  the second term will be  small
compared to the first. [For example, if the potential has the form  
$\bar{V}_R(\bar{\phi}) \propto \exp(- \bar{\phi}/\phi_0)$, the appropriate form 
of the second term would be $f(R) \exp(-2 t)$]. 
We shall now show that for
a particular choice of $A(R)$, we shall be able to produce expected behaviour of
the equation of state at large as well as galactic scales.

At small scales, evolution could have proceeded to the asymptotic limit so
that $V\to 0, \dphi \to1$ and a dust like component prevails, which would
require  $A(R) \to 1$. Then 
we get for the average field
\bear
\sqrt{1 - \del^i \phi \del_i \phi} &\approx& 
\sqrt{1 - \dot{\phi}^2 - \del^{\mu} \phi \del_{\mu} \phi}
\nline
&=& \f{\sqrt{6 f(R)}}{t^2} + {\cal O}\left({1 \over t^4}\right).
\ear
Thus, at these scales, in the limit $t \to \infty$, we have
\be
\rho_{\rm DM} \approx \f{V_0 \phi_0^2}{\sqrt{6 f(R)}}; ~~
\rho_V  \approx 0.
\e
This means that the dynamics at galactic scales is dominated by the 
pressure-less component, whose the energy density is independent of time 
\cite{asen,frolov}. 
This
resembles the non-interacting dark matter, which can cluster and 
is crucial for structure
formation in the universe. The  time dependence 
of the second term in the 
right hand side of (\ref{phi_t}) was chosen so as to make the energy 
density $\rho_{\rm DM}$ 
independent of time. In a more general scenario, this energy density will be
time dependent and will represent the standard growth of structure in the
dust-like component in an expanding universe.

Let us now turn into large scales to study the expansion of the universe. 
Since the fluctuations are likely to decrease with the averaging scales,
$\phi(R)$ will be a decreasing function of $R$ and we expect  $A(R)$  to
have a value less than unity at large scales.  Taking  $\dot{\phi}(R) = A(R) =
{\rm constant}$,  and $V = V_0 \phi_0^2/A(R)^2 t^2$ one can find consistent set
of solutions for  an $\Omega=1$ FRW model with a power law expansion $a(t)
\propto t^n$, where (see \cite{paddy} for a description of this solution): 
\be
\phi(t) = \sqrt{2\over 3n} ~ t + b_0; \quad
V(t) = {3n^2\over 8\pi G}\sqrt{1- {2\over 3n}} ~ {1\over t^2}
\label{paddy_soln}
\e
with $b_0$ being some constant. Our model reproduces the correct behaviour expected at large scales, 
provided we identify
\be
A(R) = \sqrt{2\over 3n}, \quad
V_0 \phi_0^2 = \f{n}{4 \pi G}\sqrt{1- {2\over 3n}}.
\e
Thus the average value of $\phi$ being different at different scales allows the 
possibility of the same scalar field exhibiting different equations of state
at different scales. The rate of expansion of the universe is essentially
determined by $A(R)$ at the larger scales. 

Since the same physical entity provides the dark matter at all scales
in this scenario, one certainly expects a relation between 
the energy densities contributed
by dark matter ($\Omega_m$) and dark energy ($\Omega_V$). 
In our model, the energy densities for the two components are given by
\bear
\rho_{\rm DM} &\approx& \f{V_0 \phi_0^2}{\sqrt{1 - A(R)^2}} \f{1}{t^2} 
= \f{n}{4 \pi G} \f{1}{t^2}, \nline
\rho_V  &\approx& \f{V_0 \phi_0^2 \sqrt{1 - A(R)^2}}{A(R)^2 t^2}
= \f{3 n^2}{8 \pi G} \left(1 - \f{2}{3 n} \right) \f{1}{t^2}.
\ear
(It may be necessary 
to choose the value of $V_0\phi_0^2$ in a particular range 
 to match the values of 
the energy densities we observe 
today. This could be considered a fine tuning of the parameters, which we need to resort to at this stage in the absence of a more fundamental understanding of the scalar field. It is no worse or better than the fine tuning which is required in any other model for dark energy.) 
However,
the ratio of the energy densities $\rho_V /\rho_{\rm DM}$ is 
independent of time, and 
is related to the mean value of the scalar field at large scales by
\begin{equation}
\f{\rho_V}{\rho_{\rm DM}} = \f{1}{A(R)^2} - 1.
\end{equation}
In fact, a similar equation holds for the ratio of the two components
at all scales. As one proceeds from smaller to larger scales,
the dark matter contribution decreases and the dark energy contribution
increases. 

This result can be converted into a clear prediction for cosmology
by expressing the above equation in terms of the rate of expansion $n$: 
\be
n = \f{2}{3} \left(1 + \f{\rho_V }{\rho_{\rm DM}}\right).
\e
For the values accepted at present $\rho_V /\rho_{\rm DM} \approx 2$, 
we get $a(t) \propto t^2$.
Such a rate of growth is consistent with supernova observations.
(The age of the universe in any accelerating model [with $\Omega_{tot}=1,
a(t)\propto t^n, n>1$] will be $t_0\approx n/H_0$, which is higher than the conventional models with 
$t_0\approx 1/H_0$. Any model which agrees with the SN observations and has entered an accelerating phase in the recent past will have this feature and our model with $n\approx 2$ is no different.) 
This relation between (i)~the amounts of 
dark matter and dark energy present in the universe and (ii)~the 
expansion rate is potentially testable by observations. 
It may be stressed that in our model, the evolution of the single scalar field governs the time dependence of 
{\it both} $\rho_{DM}$ and $\rho_V$. This is equivalent to saying that there is interaction and energy exchange between the two components and
the energy is \emph{not} conserved locally for the 
dark matter and dark energy components separately (which would 
imply $\rho_{\rm DM} \propto a^{-3}$ and $\rho_V = $ constant). 

Incidentally, it may be possible to put constraints on $n$ from CMB
observations as well. The pressure term in the linear perturbation equation in
this model has a factor $(1-\dot\phi^2)k^2$ where $k$ is the wave number
\cite{frolov}.  For the solution  (\ref{paddy_soln}),  this factor is
$[1-(2/3n)]k^2$ and the standard results can be used with a rescaling of $k$.
But since the angular scales of features in CMB anisotropy depends on this
rescaling, it will lead to an $n$ dependent rescaling of Doppler peaks etc. 
\cite{paddyone}. Hence, CMB observations can provide another constraint on
$n$. 

The really serious test of the model will arise from the 
non-linear small scale dynamics of the
clustering and galaxy formation scenarios. This is a hard problem which we have
not studied  in this paper; instead we have introduced an ansatz
for the form of  scalar field at different scales by hand. It is necessary to
investigate this model further and show that the basic ansatz is correct and
the details do not run into any contradiction. While this remains to  be done,
we consider it very attractive that the single entity can possibly exhibit
different equations of state at different scales in the universe. Such a
scenario has nuances (for example, for  CMB observations \cite{paddyone}) which
have not been explored in conventional cosmology before. 

We thank Ashoke Sen for useful discussions. T.R.C. is supported by the 
University Grants Commission, India.


\begin{thebibliography}{99}

\bibitem{sarkar}
S. Sarkar, in {\it EPS International Conference on High 
Energy Physics, Budapest}, 2001 (D. Horvath, P. Levai, A. Patkos, eds.), 
JHEP Proceedings Section,
PrHEP-hep2001/299; 
B. Leibundgut, {\it Ann. Rev. Astron. \& Astrophys.} {\bf 39}, 67 (2001).

\bibitem{phiindustry}
B. Ratra, and P.J.E. Peebles, {\it Phys.~Rev.~D~} {\bf 37}, 3406 (1988);
C. Wetterich, Nuclear Physics B {\bf 302}, 668 (1988);
P.G. Ferreira and M. Joyce, {\it Phys.~Rev.~D~} {\bf 58}, 023503 (1998);
J. Frieman, C.T. Hill, A. Stebbins, and I. Waga, (1995) 
{\it Phys.~Rev. Lett.~} {\bf 75}, 2077;
P. Brax and J. Martin {\it Phys.~Lett.~B~} {\bf 468}, 40 (1999);
P. Brax and J. Martin {\it Phys.~Rev.~D~} {\bf 61}, 103502 (2000);
V. Sahni and A. A. Starobinsky, IJMP {\bf 9}, 373 (2000);
  L.A. Ure\~{n}a-L\'{o}pez and T. Matos, {\it Phys.~Rev.~D~}
{\bf 62}, 081302 (2000);
T. Barreiro, E.J. Copeland and N.J. Nunes {\it Phys.~Rev.~D~} {\bf 61}, 
127301 (2000);
I. Zlatev, L, Wang and P.J. Steinhardt {\it Phys.~Rev. Lett.~} {\bf 82}, 
896 (1999);
A. Albrecht and C. Skordis {\it Phys.~Rev. Lett.~} {\bf 84}, 2076 (2000);
N. Bilic, G. B. Tupper and R. D. Viollier, {\it Phys.~Lett.~B} 
{\bf 535}, 17 (2002).


\bibitem{asen} 
A. Sen,  arXiv: hep-th/0204143; arXiv: hep-th/0203211; arXiv: hep-th/0203265 
and references cited therein.

\bibitem{comment}
This interpretation may be of some historic/pedagogical value. Generalizing the
non-relativistic  particle Lagrangian $L_{NR}=(1/2)\dot q^2-V(q)$ by changing
$q(t)$ to  field $\phi(t,{\bf x})$ will lead to standard scalar field theory
with a potential $V(\phi)$, while generalizing the relativistic particle
Lagrangian leads to the theory we are studying in the paper. Historically, one
proceeded from nonrelativistic classical mechanics to nonrelativistic quantum
mechanics and attempted to generalize the Schrodinger wave equation to
relativistic wave equations.  Instead, if one had proceeded from 
nonrelativistic classical mechanics to relativistic classical mechanics and
upgraded the $q$ to a field, one would have naturally led to this Lagrangian.
We do not know whether such an attempt was ever made in the early days of
quantum field theory. This gives another motivation to study such a scalar
field independent of its string-theoretic origin. 


\bibitem{frolov}
T. Padmanabhan, unpublished; 
A. Frolov, L. Kofman and A. Starobinsky, arXiv:hep-th/0204187;
G. Shiu and I. Wasserman, arXiv:hep-th/0205003.

\bibitem{boost}
A.Sen, hep-th/0204143 [see the discussion after equation(25)].

\bibitem{paddysfu}
T. Padmanabhan, 1993, {\it Structure formation in the universe} (Cambridge
University Press).

\bibitem{paddy}
T. Padmanabhan, arXiv:hep-th/0204150.

\bibitem{gibb} 
G.W. Gibbons, arXiv:hep-th/0204008; 
M. Fairbairn and M.H. Tytgat, arXiv:hep-th/0204070; 
S. Mukohyama, arXiv:hep-th/0204084.
D.~Choudhury, D.~Ghoshal, D.~P.~Jatkar and S.~Panda,
arXiv:hep-th/0204204.

\bibitem{earl}
C.~Acatrinei and C.~Sochichiu, arXiv:hep-th/0104263;
S.~H.~Alexander, Phys. \ Rev. \ D {\bf 65}, 023507 (2002)
[arXiv:hep-th/0105032];
A.~Mazumdar, S.~Panda and A.~Perez-Lorenzana,
Nucl.\ Phys.\ B {\bf 614}, 101 (2001)
[arXiv:hep-ph/0107058];
S.~Sarangi and S.~H.~Tye,
arXiv:hep-th/0204074.

\bibitem{fein}
A.~Feinstein,
arXiv:hep-th/0204140;

\bibitem{paddyone}
T. Padmanabhan, work in progress.




\end{thebibliography}
\end{document}